\documentclass[preprintnumbers,amsmath,amssymb,aps]{revtex4}

\begin{document}

\title{The volume enclosed by an $n$-dimensional Lam\'e curve}
\author{Ra{\'u}l Toral}

\affiliation{IFISC, Instituto de F{\'\i}sica Interdisciplinar y Sistemas Complejos, CSIC-UIB,  Campus UIB, E-07122 Palma de Mallorca, Spain  }

\date{\today}

\begin{abstract}

We compute the volume of the body enclosed by the $n$-dimensional Lam\'e curve defined by $\sum_{i=1}^nx_i^b= E$.
\end{abstract}

\maketitle

A recent paper \cite{LSC} derives asymptotic expressions for the volume of the n-dimensional body defined by  $0\le\sum_{i=1}^nx_i^b\le E$ where $b>0$. This is the body enclosed by a Lam\'e curve in $n$ dimensions.  Here we compute exactly this volume by using a straightforward modification of the calculation that gives the volume of the n-dimensional sphere, the case $b=2$, see \cite{Pathria}.

If we write $E=R^b$, the volume $V_n(R)$ is 
\begin{equation}
\begin{array}{ll}\displaystyle V_n(R)=&\displaystyle \int dx_1\cdots \int dx_n\,\,\,.\\&_{0\le\sum_{i=1}^nx_i^b\le R^b}\end{array}
\end{equation}
By dimensional analysis $V_n(R)=C_nR^n$. We next compute the integral
\begin{equation}
\int_0^{\infty}dx_1\cdots\int_0^{\infty}dx_n\exp[-(x_1^b\cdots+x_n^b)]=\left[\int_0^{\infty}dx\exp[-x^b]\right]^n=\left[\Gamma\left(1+\frac{1}{b}\right)\right]^n,
\end{equation}
by using the change of variables $r=(x_1^b\cdots+x_n^b)^{1/b}$ and the volume element $dV_n(r)=nC_nr^{n-1}dr$ as
\begin{equation}
\int_0^{\infty} dV_n(r)\exp[-r^b]=C_n\Gamma\left(1+\frac{n}{b}\right).
\end{equation}
Equaling these two expressions we get:
\begin{equation}
V_n(R)=\frac{\left[\Gamma\left(1+\frac{1}{b}\right)\right]^n}{\Gamma\left(1+\frac{n}{b}\right)}R^n.
\end{equation}
Which is the desired result. It coincides with the asymptotic limit $n\to\infty$ found in \cite{LSC} although, as shown here, the expression is valid for any value of $n$.

\end{document}